\documentclass[proof]{WileyASNA-v1}
\articletype{Article Type}%
\usepackage{filecontents}

\received{16 February 2021}
\revised{X xxxx 2021}
\accepted{X xxxx 2021}

\newcommand{\gcc}{~g.cm$^{-3}$}	

\newcommand{\rhopdf}{$\rho$-PDF}
\newcommand{\Npdf}{$N$-PDF}	

\newcommand{\Plfit}{{\sc Plfit}}  
\newcommand{\bPlfit}{{\sc bPlfit}}

\begin{document}

\title{Extraction of a second power-law tail of the density distribution in simulated clouds}

\author[1]{L. Marinkova*}

\author[1,2]{T. Veltchev}

\author[3]{Ph. Girichidis}

\author[4]{S. Donkov}

\authormark{MARINKOVA \textsc{et al}}

\address[1]{\orgdiv{Faculty of Physics}, \orgname{University of Sofia}, \orgaddress{\state{5 James Bourchier Blvd., 1164 Sofia}, \country{Bulgaria}}}

\address[2]{\orgdiv{Institut f\"ur Theoretische Astrophysik}, \orgname{Universit\"at Heidelberg, Zentrum f\"ur Astronomie}, \orgaddress{\state{Albert-Ueberle-Str. 2, 69120 Heidelberg}, \country{Germany}}}

\address[3]{\orgdiv{Leibniz-Institut f\"ur Astrophysik Potsdam}, \orgname{(AIP)}, \orgaddress{\state{An der Sternwarte 16, 14482 Potsdam}, \country{Germany}}}

\address[4]{\orgdiv{Institute of Astronomy and National Astronomical Observatory}, \orgname{Bulgarian Academy of Sciences}, \orgaddress{\state{72 Tsarigradsko Shosse Blvd., BG-1785, Sofia}, \country{Bulgaria}}}

\corres{*Corresponding author, \email{ln@phys.uni-sofia.bg}}

\abstract{The emergence and development of a power-law tail (PLT) at the high-density end of the observed column-density distribution is thought to be indicative for advanced evolution of star-forming molecular clouds. As shown from many numerical simulations, it corresponds to a morphologically analogous evolution of the mass-density distribution (\rhopdf). The latter may display also a second, shallower PLT at the stage of collapse of the first formed protostellar cores. It is difficult to estimate the parameters of a possible second PLT due to resolution constraints. To address the issue, we extend the method for the extraction of single PLTs from arbitrary density distributions suggested by \citet{VeltchevEA2019} to detect a second PLT. The technique is elaborated through tests on an analytic \rhopdf{} and applied to a set of hydrodynamical high-resolution simulations of isothermal self-gravitating clouds. In all but one case two PLTs were detected -- the first slope is always steeper and the second one is typically $\partial \ln V /\ln \rho \sim -1$. These results are in a good agreement with numerical and theoretical works and do suggest that the technique extracts correctly double PLTs from smooth PDFs.}


\keywords{ISM: clouds, ISM: structure, methods: statistical}

\maketitle


\section{Introduction}
\label{Introduction}

The birthplaces of stars are located in molecular clouds (MCs). At the initial stage of cloud formation, interstellar warm atomic gas is compressed by supersonic flows and then cools down rapidly due to non-linear thermal instabilities. The emerging MCs consist mostly of molecular hydrogen; they are turbulent, characterized by very low temperatures ($T \sim 10 - 30$~K) and an isothermal equation of state \citep[see][for a review]{BPea2020}. They are not in equilibrium and evolve towards equipartition between the turbulent and the gravitational energy. Stars begin to form at a later evolutionary stage of the cloud when self-gravity takes slowly over and local sites of gravitational collapse emerge \citep{VSea2007}. 

Therefore the study of star formation requires an understanding of the morphological and kinematical evolution of MCs. 
This evolution could be traced, e.g., through the physical parameters of cloud fragments (clumps, cores, filaments) and their scaling relations \citep{Larson1981, SolomonEA1987, HeyerEA2009, VeltchevEA2018} or by investigation of the general cloud structure in terms of abstract scales \citep{StutzkiEA1998, HeyerBrunt2004, VDK2016, DVK2017, DibEA2020}. 

An often used approach for investigating the physics of star-forming regions is the analysis of the probability density function (PDF) of the mass density (\rhopdf) and the column density (\Npdf). In isothermal, non-gravitating fluids (like at the early evolutionary stage of MCs) with well developed supersonic turbulence the \rhopdf{} has been shown to be mostly lognormal \citet{VS1994} which was later confirmed from many numerical studies \citep[e.g.,][]{PadoanEA1997, Klessen2000, LiEA2003, FederrathEA2010}. If one considers a gas volume with mean density $\rho_0$, a purely lognormal \rhopdf{} is described by the formula
\begin{equation}
\label{Equation: lognormal function}
p(s)\,ds = \frac{1}{\sqrt{2\pi\sigma^2}} \exp \Bigg[ -\frac{1}{2} \bigg( \frac{s-s_{\rm max}}{\sigma} \bigg)^2  \Bigg]\,ds~, 
\end{equation}
where $s=\log(\rho/\rho_0)$ is the logdensity and $s_{\rm max}=-0.5\sigma^2$ corresponds to its peak, with $\sigma$ being the standard deviation. {\bf As} self-gravity becomes important in the energy balance of the cloud, a power-law tail (PLT) with negative slope $q$ forms at the high-density end of the \rhopdf{} due to emerging sites of local collapse \citep{Klessen2000, KNW2011, CollinsEA2012, FederrathKlessen2012, FederrathKlessen2013, BurkhartEA2017}. The functional form of a developed PLT is:
\begin{equation}
\label{Equation: PLT}
p(s) = A\exp(qs) = A(\rho/\rho_0)^q~~,~~~~s\ge s_{\rm PLT}~,
\end{equation}
where $A$ is a constant and $s_{\rm PLT}$ is the deviation point (hereafter, DP) from the main, quasi-lognormal part of the \rhopdf{}. At the stage of ongoing local collapses in the cloud, the slope of the PLT becomes slowly shallower and tends toward some constant value while the DP shifts toward lower densities \citep{GirichidisEA2014, VeltchevEA2019}.

The \Npdf{} can be directly derived from observations. In star-forming clouds, it turns out to be morphologically similar to the \rhopdf{} and consists of a lognormal main part including the peak and a PLT at the high-density end \citep{SchneiderEA2013, SchneiderEA2015a, PokhrelEA2016, SchneiderEA2021}. This is consistent with the results from simulations of self-gravitating, contracting clouds \citep{BPea2011, FederrathKlessen2013, KoertgenEA2019}. To sum up, the formation and development of PLTs both in \rhopdf{} and \Npdf{} in evolved star-forming MCs is an established phenomenon supported from observational and numerical studies. 

The analysis of the \rhopdf{} at advanced evolutionary stages derived from the high-resolution simulation (reaching down to AU scales) of star-forming clouds by \citet[][hereafter, KNW11]{KNW2011} hints at the emergence of a {\it second} PLT. Those authors found that an extended PLT with $q\sim-1.7$ is well developed at timescales about 40\% of the free-fall time, departing from the initial lognormal distribution at $s_{\rm PLT} \sim1$ and spanning more than 6 orders of magnitude in density (see Fig. 1 in KNW11). An even shallower tail is detected at very high densities ($s\gtrsim 7$) and with slope $q\simeq -1$; the authors interprete it as an indication of mass pile-up due to an additional support against gravity from the conservation of angular momentum. A second PLT in the \Npdf{} has been derived also from {\it Herschel} maps of a dozen Galactic star-forming regions \citep{SchneiderEA2015b, SchneiderEA2021}.

The issue with a possible second PLT was recently addressed in theoretical works. Modelling a self-gravitating, isothermal turbulent cloud with steady-state accretion and small homogeneous core \citet{DonkovStefanov2019} obtained two different slopes for the PLT: $q=-3/2$ far from the core and a free-fall solution $q=-2$ near to the core. The latter corresponds to a steeper second PLT, in contrast to the finding of KNW11. The model was further elaborated as the cloud thermodynamics was modified -- the gas is considered isothermal far from the cloud core but obeys an equation of state of a `hard polytrope' near to it where the core's gravitational potential dominates in the energy balance \citep{DSVK2021}. One of the obtained solutions for the density profile yields $q=-1$; this is the case with polytropic exponent $4/3$ which allows for energy balance of thermal pressure against both the gravity of the core and the ram pressure of the infalling outer shells. Thus the result of KNW11 was reproduced within a different physical framework.

The detections of a second PLT in $\rho$-/\Npdf s and the attempts to interpret the underlying physics prompt for a reliable method for PLT extraction. The widely used techniques found in the literature suffer from two major flaws: i) they depend on the chosen bin size; and, ii) they adopt an additional assumption about the non-powerlaw part of the PDF (e.g., a lognormal shape). To solve the issue, \citet[][hereafter, V19]{VeltchevEA2019} suggested a method for a single-PLT extraction which assesses the power-law part (if existing) of an arbitrary distribution as the derived PLT parameters are averaged over a set of PDF realizations with different total number of bins. Being based on the statistical approach \bPlfit{} \citep{VirkarClauset2014}, it is labeled {\it adapted} \bPlfit{} {\it method}. V19 demonstrated its effectiveness by extraction of single PLTs from \rhopdf s of simulated evolving MCs as well from \Npdf s of two star-forming regions in the Milky Way observed by {\it Herschel}. In this paper we extend further the adapted \bPlfit{} method to enable extraction of a possible second PLT from \rhopdf/\Npdf{} with a detected first PLT. The suggested technique preserves the abovementioned advantages of this method; it merely introduces and varies a cutoff in the lower-density part of the PDF. We demonstrate its applicability on \rhopdf s of simulated clouds at the stage of star formation.

The paper is structured as follows: In Section \ref{Review_of_the_adapted_bPlfit} we briefly review the adapted \bPlfit{} method. The technique which allows for detection of a second PLT is presented and substantiated in Section \ref{Elaboration_of_the_adapted_bPlfit}. Section \ref{Test of the technique} provides information on the used simulations of self-gravitating MCs and the results from application of the suggested technique to \rhopdf s derived at the final timestep of each run. We conclude with a summary in Section \ref{Conclusions}.   

\section{The adapted \bPlfit{} method}
\label{Review_of_the_adapted_bPlfit}

A full description and substantiation of the adapted \bPlfit{} method is given in V19; here we recall only its main features.  The technique can be applied to binned distributions constructed with an arbitrary binning scheme (linear, logarithmic etc.). It is appropriate for work with large datasets from numerical simulations and high-resolution imaging of MCs. The power-law fit of a distribution or part of it is derived by the use of Kolmogorov-Smirnov (KS) goodness-of-fit statistics as described in \citet{ClausetEA2009}. This procedure (called \Plfit) does not rule out that other, non-power-law, functions might better fit the observed distribution -- it simply derives the range and the slope of the best {\it possible} power-law fit. 

The adapted \bPlfit{} method derives an {\it average} slope and DP of the PLT as the total number of bins $k$ is varied. The output of the \bPlfit{} procedure applied to a single binned distribution depends on $k$. For some choices of $k$, false extractions can occur due to incompleteness of the data -- the suggested optimal PL fit is based on a few bins at the high-density end. This turned out to be a peculiarity of the \bPlfit{} procedure itself and cannot be overcome by, e.g., introducing upper cut-off of the PDF. Instead, the approach of V19 is to consider only PDFs constructed with such choices of $k$ for which \bPlfit{} yields PLTs with large spans, above some plausible value (typically, 5 bins). Variations of $k$ in a large range from a few dozens to hundreds provides a rich sample of PDFs with plausible PLT extractions; the derived slopes and DPs are used to calculate average PLT parameters. The latter are not sensitive to spikes and other local features of the PDF tail.

\begin{figure}[t]
\centerline{\includegraphics[width=88mm]{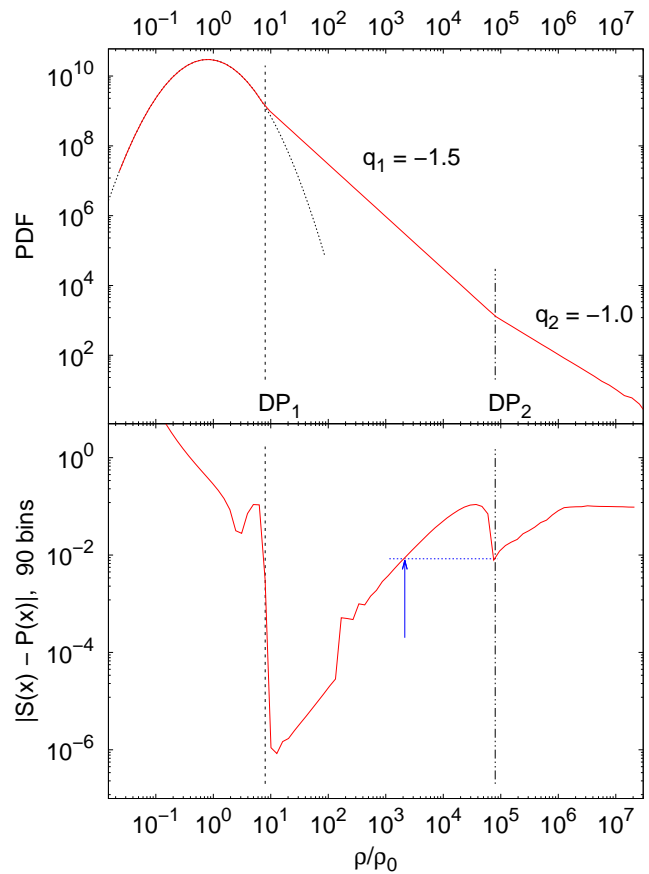}}
\caption{Illustration of the suggested method for extraction of two PLTs as applied to analytic binned PDF (top panel; red) which consists of main lognormal part (dotted line) and two PLTs (with slopes and DPs shown). The difference between the analytic CDF and the CDF of the fit is plotted in the bottom panel (red) with DPs denoted in the same way. The minimal lower density cutoff which allows for detection of the second PLT (see text) is indicated with blue arrow.}
\label{fig_analytic_PDFs_two_tails}
\end{figure}

\section{Approach for the extraction of a second PLT}
\label{Elaboration_of_the_adapted_bPlfit}

The adapted \bPlfit{} method can be elaborated further to detect a second PLT (if present). The \Plfit{} procedure searches for a PLT of the considered PDF by use of the KS statistic for a given lower cutoff $x_{\rm min}$:
\begin{equation}  \label{eq_KS_statistic}
 D = \max_{x_i \ge x_{\min}} |S(x_i) - P(x_i)|
\end{equation}
where $S(x_i)$ is the cumulative distribution function (CDF) of the data and $P(x)$ is the CDF of the best-fitting power-law model in the range $x_i \ge x_{\rm min}$ \citep{ClausetEA2009}. The deeper a minimum in $|S(x_i) - P(x_i)|$ the more likely it is to detect a transition to a PLT at $\sim x_i$. The value $x_i \ge x_{\min}$ which minimizes $D$ and the corresponding power-law index are selected as DP and slope of the PLT, respectively.

If no lower cutoff is introduced, $x_{\min}$ is simply the lower limit of the data set (in our case, the minimal logdensity) -- V19 extracted single PLTs from numerical and observational PDFs in this way. Gradual increase of $x_{\min}$ constrains the considered data set and, hence, the set of values $|S(x_i) - P(x_i)|$ to obtain the KS statistic. This approach could yield another optimal PLT estimated by the method. In particular, it may help to detect a second PLT corresponding to higher logdensities, for some $x_{\min}$ which exceeds the DP of the single (first) PLT. Let us illustrate this by a simple experiment.  

\begin{figure}[t]
\centerline{\includegraphics[width=88mm]{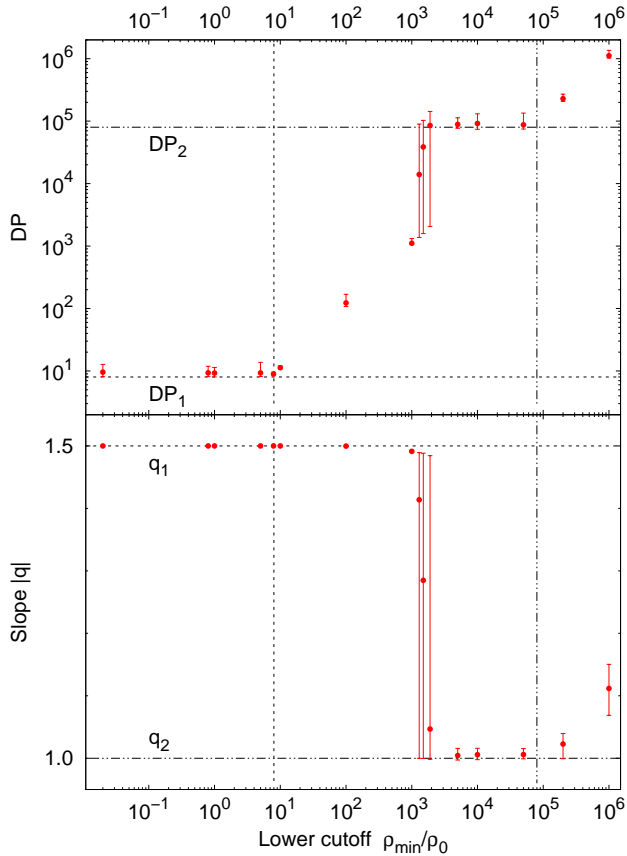}}
\caption{Dependence of the extracted PLT parameters (red symbols) on the chosen lower cutoff of the tested analytic PDF shown in Fig. \ref{fig_analytic_PDFs_two_tails} with the same notations of the PLT parameters.}
\label{fig_Cutoff-PLT_parameters_analytic}
\end{figure}

We construct an analytic PDF whose shape and parameters resemble the ones obtained in the numerical study of KNW11. It is plotted in Fig. \ref{fig_analytic_PDFs_two_tails}, top panel. The main part is lognormal while the high-density one consists of two PLTs with deviation points DP$_1$ and DP$_2$ and with slopes $q_1=-1.5$ (typical for evolved self-gravitating clouds, \citealt{GirichidisEA2014}) and $q_2=-1$ (found by KNW11). An example of the corresponding function $|S(x) - P(x)|$ for a large total number of bins (i.e. small bin size) is shown in the bottom panel of Fig. \ref{fig_analytic_PDFs_two_tails}. As expected, high values of $|S(x) - P(x)|$ are obtained in the range $x<\mathrm{DP}_1$ which defines the lognormal part of the PDF. The deviation points of the PLTs correspond to pronounced local minima of this function, with a local maximum in between. As long as $x_{\min}\equiv\rho_{\min}/\rho_0 <\mathrm{DP}_1$, the adapted \bPlfit{} will extract a single PLT with DP$=\mathrm{DP}_1$ and slope $q_1$. Choosing lower cutoffs $x_{\min}\gtrsim \mathrm{DP}_1$ still yields a single PLT with gradually changing parameters. The second PLT with DP=DP$_2$ is detected at a cutoff with $|S(x_{\min}) - P(x_{\min})|\gtrsim |S({\rm DP}_2) - P({\rm DP}_2)|$ (blue arrow and dotted line in Fig. \ref{fig_analytic_PDFs_two_tails}, bottom) -- then the procedure selects $x_i=\mathrm{DP}_2$ since this choice corresponds to the absolute minimum in the considered dataset with $x\ge x_{\min}$.

Fig. \ref{fig_Cutoff-PLT_parameters_analytic} illustrates how the chosen lower cutoffs of the analytic PDF from Fig. \ref{fig_analytic_PDFs_two_tails} affect the extracted PLT parameters (slope and DP). Note that the error bars ($\Delta q^{-}, \Delta q^{+}$) and ($\Delta {\rm DP}^{-}, \Delta {\rm DP}^{+}$) do not indicate the standard deviations but the minimal and maximal values in the sample of individual PDFs over which the average ($q$, DP) have been derived; see also V19. For $x_{\min}\le$DP$_1$, the method detects a single (i.e. the first) PLT with a very high precision. The obtained slopes practically coinside with $q_1$ and the DPs are systematically but slightly higher than DP$_{1}$ which is an effect of the binning. Lower cutoffs in the range DP$_1\le x_{\min}\lesssim10^3$ still yield a single PLT with slopes $\approx q_1$ while the values of DP increase gradually and are proportional to the chosen $x_{\min}$. With choices of $x_{\min}$ close to $2\times10^3$ (marked with arrow in Fig. \ref{fig_analytic_PDFs_two_tails}, bottom) the method starts to detect the second PLT, for some particular bin sizes -- this leads to large uncertainties of the obtained slope which cover the range $[q_1,~q_2]$. However, the transition to the detection of a single second PLT is almost abrupt (Fig. \ref{fig_Cutoff-PLT_parameters_analytic}); as soon as $x_{\min}$ becomes larger than the critical value $2\times10^3$ for {\it all} considered bin sizes\footnote{That is, for all considered individual binned PDFs with plausible PLT extractions.}, the parameters of the second PLT are reproduced with small uncertainties. But for choices $x_{\min}>$DP$_2$ their values deviate from ($q_2$, DP$_2$) increasingly, due to effects of insufficient statistics at the high-density end of the \rhopdf. 

Evidently, the suggested modification of the adapted \bPlfit{} method extracts successfully a second PLT from smooth PDFs. In the next Section we apply it to data from simulated self-gravitating clouds.

\section{Application on numerical data} \label{Test of the technique}
\subsection{Hydrodynamical simulations}
\label{Data}

We make use of data from six runs of isothermal hydrodynamical simulations of self-gravitating turbulent clouds at scales of typical large clumps ($0.5$ pc) in MCs. The set of simulations is called HRIGT (High-Resolution Isothermal Gravo-Turbulent) and is described in more detail in V19. The complexity in physical modeling is reduced in favor of higher resolution and significantly higher adaptive refinement (from $256^3$ up to $32768^3$ cells) which yields resolutions down to $\sim3$ au in high-density zones. The gas is isothermal ($T=10$ K) and uniformly distributed at the initial point in time. The total mass in the box is chosen to be 32 or 354 Jeans masses ($M_{\rm J,0}$) in the different runs which corresponds to the moderately and strongly self-gravitating regime, respectively. Lagrangian sink particles are defined at densities above the threshold $2\times10^{-13}$~\gcc. The turbulent flows are slightly supersonic, with Mach number 2, and their initial velocities are constructed in Fourier space with a peak of the power spectrum at scales corresponding to half of the box size. Three types of velocity field modes are varied: purely compressive (c), purely solenoidal (s) and naturally mixed (m) velocities \citep{FKS2008}. The total run time of the selected HRIGT simulations varies from $0.4$ up to $2.5$ free-fall times. Some parameters of the numerical runs are given in Table \ref{table_HRIGT_runs}, columns 1-6.

In view of their characteristics and physics, the selected HRIGT runs are appropriate to test our technique for the extraction of a second PLT. The box size and its Jeans content combined with low Mach number reflect the condinitions in dense self-gravitating clouds while the high adaptive refinement and the achieved advanced evolutionary stages allow to resolve a possible second PLT. Below we present the results from such tests.

 \begin{table*}
\caption{HRIGT runs selected to elaborate the method for PLT extraction and the obtained PLT parameters. Notation: v1, v2, v3 -- chosen velocity field; vfmode -- velocity field mode; mix(ed), com(pressive), sol(enoidal); $M_{{\rm J},\,0}$ -- initial Jeans mass in the box; $\tau_{\rm ff}$ -- free-fall time derived from the mean density $\langle\rho\rangle$; $t_{\rm sim}$ -- run duration; SPT -- sink particles' threshold.}
 \label{table_HRIGT_runs} 
 \begin{center}
 \begin{tabular}{rc@{~~}c@{~}c@{~~~}c@{~~}c@{~~~~~}c@{~~~}c@{~~~~~}c@{~~~}cr}
 \hline 
 \hline 
 Name & Total mass  & vfmode & $\langle\rho\rangle$ & $\tau_{\rm ff}(\langle\rho\rangle)$ & $t_\mathrm{sim}$ & $|q_1|$ & DP$_1$ & $|q_2|$ & DP$_2$ &  SPT/DP$_2$\\ 
 ~     & [$M_{{\rm J},\,0}$] & ~ & [$\rm g.\rm cm^{-3}$] & [ kyr ] & [ $\tau_{\rm ff}$ ] \\
 \hline 
 v1m-M085 & ~32 &	mix  & ~$4.6\times10^{-20}$ & $310$ & $1.05$ & $1.60~ \scriptstyle \pm {0.030}$ & $7.6\times10^2$ & $0.91~ \scriptstyle \pm {0.020}$ & $2.5\times10^4$& $~174$ \\
 v1c-M426 & 354 &	com  & ~$2.3\times10^{-19}$ & $139$ & $0.41$ & $1.96~ \scriptstyle \pm {0.020}$ & $7.7\times10^1$ & $1.05~ \scriptstyle \pm {0.002}$ & $2.7\times10^4$& $~~32$ \\	
 v1s-M426 & 354 & 	sol  & ~$2.3\times10^{-19}$ & $139$ & $1.57$ & $2.27~ \scriptstyle \pm {0.003}$ & $2.7\times10^1$ & $1.51~ \scriptstyle \pm {0.040}$ & $4.3\times10^2$& $2022$ \\
 v2m-M085 & ~32 &	mix  & ~$4.6\times10^{-20}$ & $310$ & $2.50$ & $1.58~ \scriptstyle \pm {0.030}$ & $1.9\times10^2$ & $0.86~ \scriptstyle \pm {0.010}$ & $6.2\times10^4$& $~~70$ \\	
 v2m-M426 & 354 &	mix  & ~$2.3\times10^{-19}$ & $139$ & $0.80$ & $2.55~ \scriptstyle \pm {0.010}$ & $3.7\times10^1$ & $1.03~ \scriptstyle \pm {0.010}$ & $1.2\times10^4$& $~~72$ \\
 v3m-M085 & ~32 &	mix  & ~$4.6\times10^{-20}$ & $310$ & $1.38$ & --  & -- & $1.13~ \scriptstyle \pm {0.060}$ & $1.3\times10^4$& $334$ \\
 \hline 
 \hline 
 \end{tabular} 
 \end{center}
 \smallskip 
 \end{table*}

\subsection{Extractions of double PLTs}
Emergence and evolution of single PLTs in the HRIGT runs was studied in V19. The initial velocity field leads to the formation of a quasi-lognormal \rhopdf{} while in all but one case a PLT can be distinguished ($|q|\lesssim4$) as late as at 80\% of the run time (see Fig. 8, left, in V19). At the final stages of the runs -- as sink particles (first stars) form -- the slopes tend towards some constant value, though with large uncertainties which indicate the emergence of a second PLT. To test the suggested technique for extraction of a second PLT we selected data boxes at the very end of the runs (given in column 6 in Table \ref{table_HRIGT_runs}) to construct the \rhopdf.

Like in V19 (see also Sect. \ref{Review_of_the_adapted_bPlfit}), the PLT parameters are derived simultaneously which circumvents the arbitrariness of the DP assessment by eye and the bias of the slope introduced in this way. The procedure of averaging of the PLT parameters over a set of PDF realizations with different total number of bins for {\it each chosen} $\rho_{\min}/\langle \rho\rangle$ removes any dependence of the assessed PLT on the bin size.  

\begin{figure*}[t]
\centerline{\includegraphics[width=16cm]{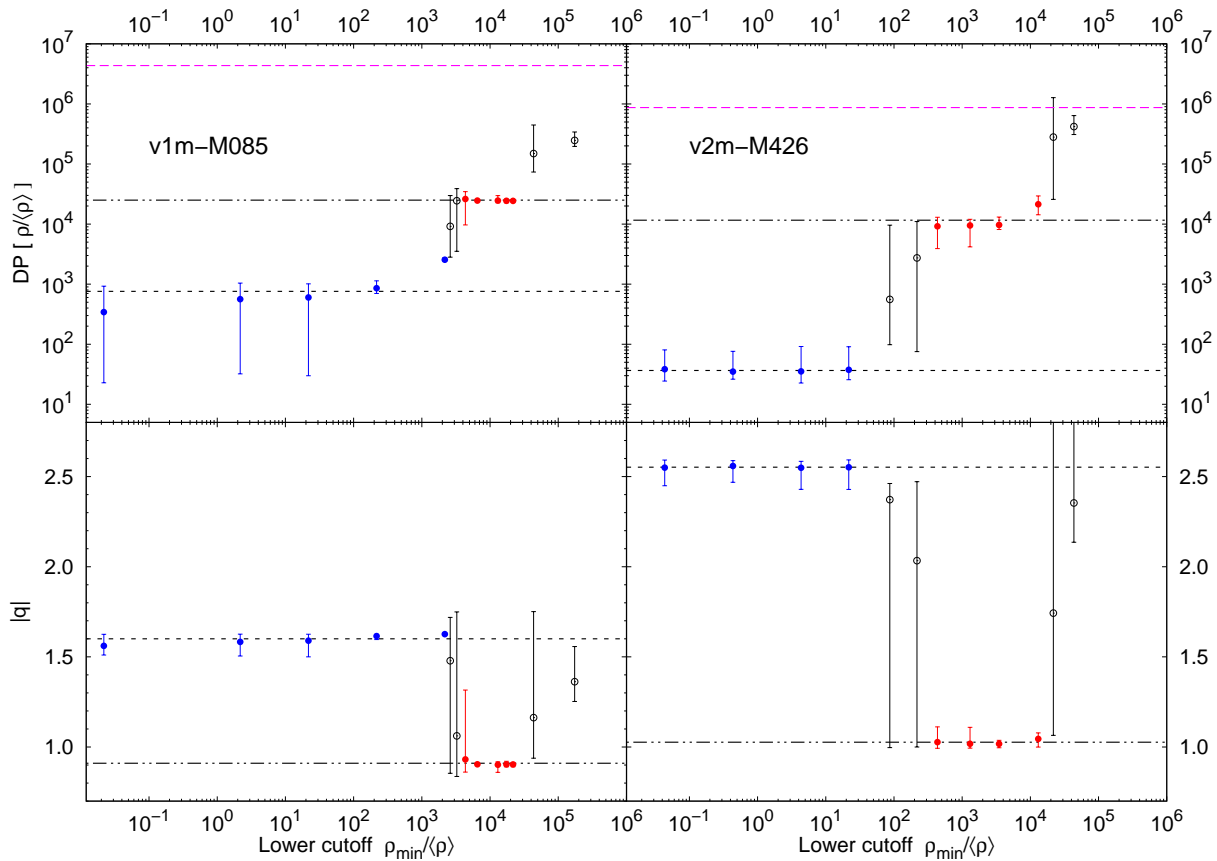}}
\caption{Dependence of the extracted PLT parameters from two HRIGT runs on the chosen lower cutoff. The points selected to assess the first (blue) and the second (red) PLT are plotted as well those (black, open) which do not meet the adopted criteria (see text). The averaged PLT parameters are denoted with black dashed (1st PLT) and dash-double dotted (2nd PLT) lines; the sink particles' threshold is shown (magenta, dashed) for comparison.}
\label{fig_Cutoff-PLT_parameters_HRIGT}
\end{figure*}

Two illustrations on how the extracted PLT parameters depend on the chosen lower cutoff $\rho_{\min}/\langle \rho\rangle$ are shown in Fig. \ref{fig_Cutoff-PLT_parameters_HRIGT}. Their behavior resembles the one of the PLT parameters extracted from the analytic PDF (cf. Fig. \ref{fig_Cutoff-PLT_parameters_analytic}). First, one clearly distinguishes two groups with very close estimates, within the typical uncertainties, of DPs and slopes. These groups are to be associated with two detected PLTs. Second, the two groups are separated by a narrow range of $\rho_{\min}/\langle \rho\rangle$. Here, the extracted PLT parameters are characterized by uncertainty ranges comparable with the difference between the mean PLT parameters of the two groups. Third, choices of $\rho_{\min}/\langle \rho\rangle$ within the presumed range of the second PLT yield essentially different estimates compared to the mean PLT parameters of the second group. The deviations grow as the chosen cutoff approaches the sink particles' threshold (SPT) in density.

This morphological analysis of the PDF features suggests that one can safely assess the average parameters of the presumed two PLTs imposing some requirements to exclude points corresponding to $\rho_{\min}/\langle \rho\rangle$ from the transition range or at the high-density PDF end. In this work we adopt rather conservative criteria regarding the plausible detectability of a second PLT. The procedure is as follows:
\begin{enumerate}
 \item {\it Initial estimates of the PLT parameters} ($\tilde{q_1}$, $\widetilde{\rm DP}_1$) and ($\tilde{q_2}$, $\widetilde{\rm DP}_2$) by eye ($q$ vs. $\rho_{\min}/\langle \rho\rangle$ and DP vs. $\rho_{\min}/\langle \rho\rangle$ diagrams). Typical standard deviations of the individual points should be taken as uncertainties of these estimates. Here we require for the extraction of two PLTs: 
 \begin{itemize}
  \item $|\tilde{q_1} - \tilde{q_2}|\ge0.4$
  \item SPT/$\widetilde{\rm DP}_2>10$, i.e. the span of the second PLT to be at least one order of magnitude.
 \end{itemize}
 \item {\it Exclusion of points that may distort the PLT estimates:}
 \begin{itemize}
  \item The transition range in densities between the presumed PLTs:  $\tilde{q_1} < q < \tilde{q_2}$, $\widetilde{\rm DP}_1$ < DP < $\widetilde{\rm DP}_2$ {\bf and} $(\Delta q^{+} + \Delta q^{-})\sim |\tilde{q_1} - \tilde{q_2}|$
  \item High-density end of the PDF: $\rho_{\min}/\langle \rho\rangle$ > $a \,\times\,\widetilde{\rm DP}_2$, where $a>1$ is a small factor depending on the typical standard deviations.
 \end{itemize}
 \item {\it Averaging of the PLT parameters ($q_1$, {\rm DP}$_1$) and ($q_2$, {\rm DP}$_2$) over the groups associated with the presumed PLTs}  
\end{enumerate}

Table \ref{table_HRIGT_runs}, columns 7-11, contains the parameters of the PLTs extracted from the selected HRIGT runs. They are juxtaposed for illustration with single PDF realizations, for a fixed large total number of bins, in \ref{Appendix}, Fig. \ref{fig_PDFs_HRIGT_runs}. In all but one case our technique extracts two PLTs which span at least one order of magnitude each. The first PLTs deviate from the main PDF part at $\rho/\langle \rho\rangle\sim 10^1-10^2$ and their slopes vary in the relatively large range $-2.6 \lesssim q_1 \lesssim -1.6$. The extracted second PLTs are more uniform in regard to their parameters, with DP of order $10^4$ and slopes close to $-1$. The obvious exception is the run v1s-M426; in this case the second PLT is steeper and also spans a much larger range in densities than the first PLT. The single PLT detected from the run v3m-M085 is most probably the {\it second} one, in view of the parameters, while the first one has been not detected due to clumpy PDF in the range $10^1 \lesssim \rho/\langle \rho\rangle \lesssim 10^4$.  

In general, in all cases with two detected PLTs the first PLT is steeper than the limit $\sim -1.5$ expected from theoretical considerations \citep{GirichidisEA2014}. This limit is approached in two runs characterized by the shallowest second PLTs as well -- this points to the late evolutionary stage of the simulated clouds. Slopes about $-1$ of the second PLTs are in general agreement with the simulation of star-forming clouds of KNW11 discussed in Sect. \ref{Introduction}. These findings suggest that our technique extracts correctly double PLTs from smooth PDFs. Its tests to column-density PDFs from observations and large-scale numerical data are subjects of further study. 

\section{Discussion} \label{Discussion}

The presented technique enables thorough analysis of the high-density end of the $\rho$-/\Npdf~in star-forming regions and thus can contribute essentially to the physical understanding of their densest substructures (clumps and cores). First, it can provide a correct slope estimate of a second PLT which is to be compared with the theoretical predictions. Recently, there are several physical explanations of a second PLT in the \rhopdf. KNW11 interpreted the second PLT with slope $q\simeq -1$ found at advanced evolutionary stage of their simulated self-gravitating clouds as an effect of rotation in emerging protostellar cores providing additional support against gravity. On the other hand, \citet{DSVK2021} argue that the transition between the first and the second PLT of the cloud's \rhopdf~may mark a change in the thermodynamical state of the gas. According to their model of self-gravitating accreting cloud, the equation of state $P_{\rm gas}\propto\rho^{\Gamma}$ changes from roughly isothermal ($\Gamma = 1$) in the outer regions to that of a `hard  polytrope' with $\Gamma>1$. The obtained solutions correspond to different density profiles, polytropic exponents and energy balance in a considered fluid element; two of them yield $q=-2$ ($1 < \Gamma < 4/3$) and $q=-1$ ($\Gamma = 4/3$) and can be considered as descriptions of consecutive stages of cloud evolution. Strong magnetic fields in the cloud which oppose gravity and, hence, slow down the accretion could also produce a second PLT in the \rhopdf~(KNW11).

Second, the suggested technique can discriminate between clear examples of observational \Npdf s with two PLTs \citep[like in][]{SchneiderEA2015b} and such where the existence of a second PLT is questionable. A case of special interest is a \Npdf{} with a {\it steeper} second PLT. If it can be demonstrated from observations, that would pose a new challenge to the theory of evolving self-gravitating clouds.

\section{Conclusions} \label{Conclusions}

The technique we present for the extraction of a second power-law tail (PLT) of the density distribution (\rhopdf) or of the column-density distribution (\Npdf) in star-forming clouds is an extension of the adapted \bPlfit{} method \citep{VeltchevEA2019}. Its idea is based on the dependence of the extracted PLT parameters on the chosen lower cutoff in density -- varying the latter allows for detection of two PLTs. We elaborated the method through tests on an analytic \rhopdf{} with main lognormal part and two PLTs. Then it was tested on data from numerical simulations of self-gravitating isothermal star-forming clouds at clump scale (0.5 pc), with high resolution and adaptive mesh refinement in zones of high density. We require that the second PLT must be at least one order of magnitude and its slope must differ at least by $0.4$~dex from the slope of the first PLT. In all but one case two PLTs were detected -- the first one (deviating from the quasi-lognormal part) is steeper and the slope of the second one is flatter than $-1.5$ and tending towards $\sim -1$. The obtained results are in a good agreement with numerical \citep{KNW2011} and theoretical \citep{GirichidisEA2014} studies. Hence we argue that the presented extension of the adapted \bPlfit{} technique can be successfully used to detect a second PLT of smooth PDFs and plausible estimates of its parameters. Its application to (column-)density distributions from high-resolution observational data and/or numerical simulations of star-forming regions can elucidate the physical conditions in their densest substructures.

\section*{Acknowledgements}
L. M. thanks to the \fundingAgency{Bulgarian National Science Fund} for providing support under grant \fundingNumber{KP-06-PM-38/6} (Fundamental research by young scientists and postdocs 2019). T.V. and S.D. acknowledge support by the \fundingAgency{Deutsche Forschungsgemeinschaft (DFG)} under grant \fundingNumber{KL 1358/20-3} and additional funding from the \fundingAgency{Ministry of Education and Science of the Republic of Bulgaria, National RI Roadmap Project} \fundingNumber{DO1-383/18.12.2020}. P.G. acknowledges funding from the European Research Council under ERC-CoG grant CRAGSMAN-646955.

\appendix

\section{PDFs and PLT parameters from HRIGT runs}
\label{Appendix}
\begin{figure*}[t]
\centerline{\includegraphics[width=1.\textwidth]{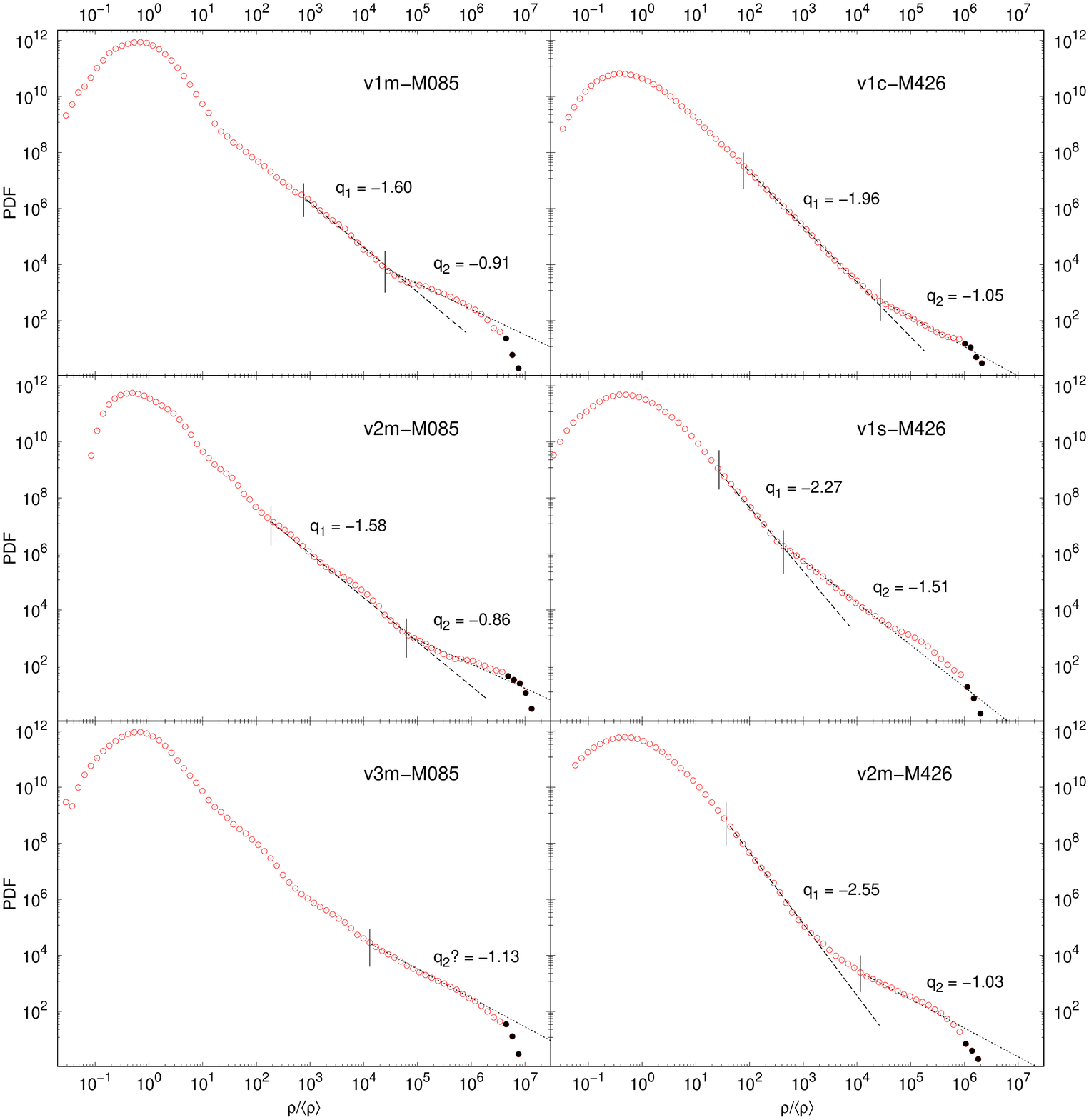}}
\caption{The extracted PLTs compared with a single realization of the PDF (for a fixed number of bins) from the selected HRIGT runs. Black dots denote neglected data at the high-density end, above the threshold density to define sink particles.}
\label{fig_PDFs_HRIGT_runs}
\end{figure*}

In Fig. \ref{fig_PDFs_HRIGT_runs} we plot single realizations of the PDFs from each HRIGT run for a fixed, large total number of bins together with the extracted PLTs whose parameters are listed in Table \ref{table_HRIGT_runs}. The main part of the PDFs is quasi-lognormal while high-density tails are, in general, very smooth which explains the small uncertainties of the derived average slopes. Note that the locations of the DPs -- especially of the first PLT -- are not necessarily where one would expect them by eye.  

\bibliography{Marinkova_et_al_2020}

\end{document}